\documentstyle[12pt,epsfig]{article}
\begin{document}
\date{}
\title{\small \phantom{.}\hfill BARI-TH 288/97 \\[1cm]
\large\bf SPONTANEOUS GENERATION  OF MAGNETIC \\ FIELD  IN 
THREE DIMENSIONAL QED \\ AT FINITE TEMPERATURE}
\author{Paolo Cea and Luigi Tedesco} 
\maketitle
\begin{center}
Dipartimento di Fisica and Sezione INFN \\
Via Amendola 173,  I-70126 Bari, Italy
\end{center}
\vspace{2cm}
\begin{abstract}
We investigate the effects of thermal fluctuations on the spontaneous magnetic 
condensate in three dimensional QED coupled with P-odd Dirac fermions. Our 
results show that the phenomenon of the spontaneous generation of the constant 
background magnetic field survives to the thermal corrections even at infinite 
temperature. We also study the thermal corrections to the fermionic condensate 
in presence of the magnetic field.
\end{abstract}
%
%
%
\newpage

Recently~\cite{CEA} it has been shown that, within the one-loop approximation,
the unique theory displaying a nontrivial ground state turns out to be the 
three-dimensional $U(1)$ gauge theory in interaction with Dirac 
fermions with
a negative P-odd mass 
term. Indeed, in that theory there is a spontaneous generation of a non-zero 
constant magnetic field. So that the model could be relevant towards the 
effective description of planar systems in condensed matter physics.

More interestingly, it is well known that Dirac fermions with a Yukawa coupling 
with a scalar field develop zero mode solutions near a domain 
wall~\cite{JACKIW}. It turns out that the zero modes behave like massless 
fermions in the two-dimensional space of the wall. In this case the formation 
of a uniform magnetic condensate and the dynamical generation of a constant 
P-odd fermion mass are energetically favorable. Thus the domain wall becomes 
ferromagnetic. Recently it has been argued  that ferromagnetic domain walls 
could be relevant for the formation of primordial magnetic 
field~\cite{IWAZAKI}. In view of this, it is important to ascertain if the 
spontaneous generation of magnetic field survives to the thermal fluctuations.

The aim of the present paper is to study the thermal corrections in the 
three-dimensional $U(1)$ gauge theory in interaction with fermionic fields with 
negative P-odd mass term. Indeed, as we said before, this theory displays a 
nontrivial ground state already at the one-loop level.

Let us consider the Hamiltonian in the temporal gauge $A_0(x)=0$:

\begin{eqnarray}
\label{eq1}
H &=& \int d^2 x \left\{  \frac {1} {2}{\vec{E}}^2(x) +\right. 
       \frac {1} {2}  B^2(x) 
      + \psi^{\dag}(x)[- i \vec{\alpha} \cdot \vec{\nabla}
      + \beta m] \psi(x)+ \nonumber \\     
  && ~~~~~~~~~~~-  \left. e \psi^{\dag}(x) \vec{\alpha} 
     \cdot \vec{A}(x) \psi(x) \right\},
\end{eqnarray}
where we follow the Bjorken and Drell notation and 
we use the two-dimensional realization of the Dirac algebra:

$$
\gamma^0=\sigma_3, \; \; \gamma^1=i \sigma_1, \;\; \gamma^2= i \sigma_2, 
$$
\begin{equation}
\label{eq2}
\{\gamma^{\mu}, \gamma^{\nu}\}=2 g^{\mu \nu}, \;\;\;  g^{\mu \nu}=diag(1,-1,-1).
\end{equation}
Previous studies~\cite{CEA,HOSOTANI} showed that if $m$ is negative, then the 
perturbative ground state is unstable toward the spontaneous formation of a 
constant background magnetic field. The magnetic instability is present even at 
the one-loop level. To see this, we use the Furry's representation. If we 
split the electromagnetic field into the fluctuation $\eta(x)$ over the 
background ${\bar {A} (x)}$:
\begin{equation}
\label{eq3}
A_k(x)={\bar{A}}_k(x) + \eta_k(x)
\end{equation}
with
\begin{equation}
\label{eq4}
{\bar{A}}_k(x)= \delta_{k 2}\, x_1\, B,
\end{equation}
then in the one-loop approximation the Hamiltonian reads ($V$ is the spatial 
volume):
\begin{eqnarray}
\label{eq5}
H_0 &=& H_{\eta}+H_D + V \frac {B^2} {2} 
 = \int  d^2 x \left \{ \frac {1} {2} {\vec{E}}^2(x) + \frac {1} {2} 
[\epsilon_{i j} \partial_i \eta_j(x)]^2 \right\}+ \nonumber \\ 
 &&  + \int d^2 x \left \{ 
\psi^{\dag}(x)[{\alpha}_k(-i\partial_k-e\bar{A}_k)
+\beta m ] \psi(x) \right \} + V \frac {B^2} {2}
\end{eqnarray}
The fermionic Hamiltonian can be diagonalized by expanding the Dirac field into 
the eigenstates of Dirac equation:

\begin{equation}
\label{eq6}
[\alpha_k(-i \partial_k-e{\bar{A}}_k)+ \beta m ]\; \psi(x) = E\; \psi(x).
\end{equation}
In the Furry's  representation we expand the fermionic operator 
$\psi(x)$ in terms of the positive and negative solutions 
$\psi^{(+)}_{np}$ and $\psi^{(-)}_{np}$
of Eq. (\ref{eq6}) . In the case of negative 
mass term $m=-|m|$  the positive solutions have eigenvalues 
$+E_n=+ \sqrt{( 2neB + m^2)}, \;\;\; n\geq 1$, while the negative ones 
have eigenvalues
$-E_n = -\sqrt{( 2neB + m^2)}, \;\;\; n \geq 0$. Therefore we write   
\begin{equation}
\label{eq7}
\psi(x)=\sum_{n=1}^{\infty} \int_{-\infty}^{+\infty} d p \; \psi^{(+)}_{n,p}(x) 
\; a_{n p}+\sum_{n=0}^{\infty} \int_{-\infty}^{+\infty} d p \; 
\psi^{(-)}_{n,p}(x) \; b^{\dag}_{n p}.
\end{equation}
Thus we get \footnote{We assume here and throughout the paper that $eB>0$.
Moreover, the  functions $\psi^{(\pm)}_{n,p}(x)$ are normalized as 
in ref~\cite{CEA}.}

\begin{equation}
\label{eq8}
H_D=\int_{-\infty}^{+\infty} d p \left[\sum_{n=1}^{\infty}
E_n \, a^{\dag}_{n p} a_{n p} + \sum_{n=0}^{\infty} E_n b^{\dag}_{np} b_{np} 
\right] - \frac {eB} {2 \pi} V \sum_{n=0}^{\infty}E_n.
\end{equation}
Observing that $\frac {eB} {2 \pi} V$ is the 
degeneracy of the Landau levels, we see that the last term in Eq. 
(\ref {eq8}) is the well known Dirac sea energy.

Now it is easy to find the vacuum energy density. Obviously we are interested 
in the difference between the vacuum energy in presence of the magnetic field
and the perturbative vacuum energy. It is a  straightforward exercise to find:

\begin{equation}
\label{eq9}
{\cal {E}}(B)= \frac {1} {V} E(B) = \frac {B^2} {2} - \frac {eB} {2 \pi} 
\sum_{n=0}^{\infty} E_n.
\end{equation}
Introducing the dimensionless variable $\lambda=\frac {e B} {m^2}$, 
the function 

\begin{equation}
\label{eq10}
g(\lambda)=\int_0^{\infty} \frac {dx} {\sqrt{\pi x}} \frac {d} {dx} \left[
\frac {e^{-\frac {x} {\lambda}}} {1-e^{-2 x}}- \frac {e^{-\frac {x} {\lambda}}}
{2 x}\right],
\end{equation}
and subtracting a term independent on the magnetic field, we recast 
the energy density in the form~\cite{CEA}:

\begin{equation}
\label{eq11}
{\cal {E}}(B)=\frac {B^2} {2}+ \frac {(eB)^{\frac {3} {2}}} {2 \pi} g(\lambda).
\end{equation}
It is easy to show that, indeed, ${\cal {E}}(B)$ displays a nontrivial negative 
minimum~\cite{CEA}. 
As a matter of fact, using the expansion~\cite{CEA}

\begin{equation}
\label{eq11bis}
g(\lambda)  \stackrel{\lambda \rightarrow  0}{\sim} 
- \frac {1} {2 \lambda^{\frac {1} {2}}} +
\frac {\lambda^{1\over 2}} {12}
\end{equation}
we see that near the origin 

\begin{equation}
\label{eq11tris}
{\cal {E}} \sim - \frac {eB} {4 \pi} |m| + \frac {1} {24 \pi} \frac {(eB)^2} 
{|m|} + \frac {B^2} {2}.
\end{equation}
The negative linear term in Eq. (\ref{eq11tris}) gives rise to the negative 
minimum in the vacuum energy density.

In the remainder of the paper we study the thermal corrections to the vacuum 
energy density. In particular we are interested in the phenomenon of the 
restoration by thermal corrections of the symmetry broken at zero temperature. 
Remarkably, it turns out that the spontaneous generation of the magnetic 
condensate survives the thermal corrections.

The relevant quantity at finite temperature is the free energy

\begin{equation}
\label{eq12}
F=-\frac {1} {\beta} \ln Z,
\end{equation}
where $\beta=\frac {1} {T}$ and $Z$ is the partition function 

\begin{equation}
\label{eq13}
Z=Tr(e^{- \beta H}).
\end{equation}
In the one-loop approximation we have

\begin{equation}
\label{eq14}
F_0=-\frac {1} {\beta}\; \ln Z_0, \;\;\;\;\; 
Z_0=Tr\left[e^{-\beta H_0}\right].
\end{equation}
In our approximation the free energy is the sum of the photonic and fermionic 
contributions. Only the latter depends on the background magnetic field $B$. 
Moreover it is a straightforward exercise to evaluate the partition function 
corresponding to the Hamiltonian $H_D$. 
Indeed, the calculation reduces to evaluating  the partition 
function of a free
relativistic fermions in presence of the  magnetic field. Taking into 
account that $V \frac {eB} { 2 \pi}$ is the degeneracy of the Landau levels we 
obtain:

\begin{equation}
\label{eq15}
{\cal F}_0(B)=\frac {F_0(B)} {V}=- \frac {2} {\beta} \frac {eB} { 2 \pi} 
\sum_{n=1}^{\infty} \ln \left(1+e^{-\beta E_n}\right) - \frac {1} {\beta} 
\frac {eB} { 2 \pi} \ln \left( 1+ e^{-\beta |m|} \right) + {\cal E}(B).
\end{equation}
The first term on the right hand of Eq. (\ref{eq15}) aries from the negative 
and positive Landau levels with $n > 0$, the second one is the zero mode 
contribution and the last term is the zero temperature vacuum energy density.

In order to evaluate the infinite sum in Eq.(\ref{eq15}) we expand the logarithm 
to obtain 

\begin{equation}
\label{eq16}
I_1(B)=- \frac {eB} {\pi \beta} \sum_{n=1}^{\infty} \ln \left(1+e^{-\beta E_n}
\right)=- \frac {eB} {\pi \beta} \sum_{n=1}^{\infty} \sum_{k=1}^{\infty} \frac 
{(-1)^{k+1}} {k} e^{-\beta E_n k}.
\end{equation}
Now using~\cite{GRADSHTEYN}

\begin{equation}
\label{eq17}
\int_0^{\infty} e^{-a x^2 - \frac {b} {x^2}} dx = \frac {1} {2} \sqrt{\frac 
{\pi} {a}} e^{- \sqrt{ a b}}, \;\;\;\; a>0, \;\;b>0
\end{equation}
we perform the sum over $n$ 
\begin{equation}
\label{eq17a}
\sum_{n=1}^{\infty} e^{-\beta k E_n}=\frac {1} {\sqrt{\pi}} \int_0^{\infty}
dx \; e^{- \frac {x^2} {4} -
\frac {\beta^2 k^2
m^2} {x^2}}  \frac {1} { e^{\frac {2 \beta^2 eB k^2} {x^2}} -1}   
\end{equation}
and obtain

\begin{equation}
\label{eq18}
I_1(B) = - \frac {eB} {\pi^{\frac {3} {2}} \beta} \sum_{k=1}^{\infty} \frac
{(-1)^{k+1}} {k} \int_0^{\infty} dx \; e^{- \frac {x^2} {4} - 
\frac {\beta^2 k^2 
m^2} {x^2}}  \frac {1} { e^{\frac {2 \beta^2 eB k^2} {x^2}} -1} \; .
\end{equation}
Subtracting the contribution at $B=0$ we get

\begin{equation}
\label{eq19}
{\tilde {I}}_1(B)= - \frac {eB} {\pi^{\frac {3} {2}} \beta} \sum_{k=1}^{\infty} 
\frac {(-1)^{k+1}} {k} \int_0^{\infty}
dx e^{- \frac {x^2} {4} - \frac {\beta^2 k^2 m^2} {x^2}} \left[ 
\frac {1} { e^{\frac {2 \beta^2 eB k^2} {x^2}} -1}  - \frac {x^2} 
{2 eB \beta^2 k^2}\right].
\end{equation}
Thus we have

\begin{equation}
\label{eq20}
{\cal {F}}_0 (B) = {\tilde {I}}_1 (B)-\frac {1} {\beta} \frac {eB} {2 \pi}\ln 
\left(1+ e^{- \beta |m|}\right)+ {\cal {E}}(B).
\end{equation}
This last expression is amenable to a numerical evaluation.
In Figure 1 we display the free energy density ${\cal {F}}_0 (B)$ (in units of 
$|m|^3$) for four different temperatures ${\hat {T}}=\frac {T} {|m|}$ and 
$\alpha = \frac {|m|} {e^2}=0.1.$ We see that by increasing the temperature
the negative minimum never 
disappears. In other words, the spontaneous 
magnetic field survives at any temperature. Similar results hold even in 
presence of the Chern-Simons term~\cite{KANEMURA}. Interesting enough, it turns 
out that the asymptotic curve at infinite temperature can be evaluated in closed 
form. Indeed, from Eq.(\ref{eq16}) we get for ${\hat {\beta}}= \beta |m|<<1$

\begin{equation}
\label{eq21}
I_1(B)= \frac {eB} {2 \pi \beta} \ln 2 + \frac {eB} {2 \pi} \sum_{n=1}^{\infty} 
E_n +{\cal {O}} (\beta^2).
\end{equation}
On the other hand, in the same approximation we have

\begin{equation}
\label{eq22}
- \frac {e B} {2 \pi \beta} \ln \left( 1+ e^{- \beta |m|}\right)= 
- \frac {e B} {2 \pi \beta} \ln 2 + \frac {eB} { 4 \pi} |m| + 
{\cal {O}} (\beta^2). 
\end{equation}
From Equations (\ref{eq9}),(\ref{eq20}),(\ref{eq21}) and (\ref{eq22}),
it follows the remarkable simple result:

\begin {equation}
\label{eq23}
{\cal {F}}_0(B)= - \frac {eB} {4 \pi} |m| + \frac {B^2} {2}+ {\cal {O}} (
\beta^2).
\end{equation}
Again we see that 
the negative minimum at finite temperature is due to the linear term  
in $|m|$. 
Note that the slope of the linear term coincides with the one at zero 
temperature. 
This can be seen clearly in Fig. 2 
where we display the free energy density without the classical energy $\frac 
{B^2} {2}$.

Equation (\ref{eq23}) displays a negative minimum at 

\begin{equation}
\label{eq24}
eB^*= \frac {e^2 |m|} {4 \pi}.
\end{equation}
The condensation energy is:

\begin{equation}
\label{eq25}
{\cal {F}}_0 (B^*)= - \frac {1} {2} eB^* |m|= - \frac {e^2 |m|^2} {32 \pi^2}.
\end{equation}
Note that the minimum Eq.(\ref{eq24}) coincides with the zero-temperature 
minimum in the "weak-coupling" region $\alpha=\frac {|m|} {e^2}>>1$~\cite{CEA}:

\begin{equation}
\label{eq26}
\frac {e B^*} { m^2} = \frac {1} {4 \pi \alpha}.
\end{equation}
Thus, even if the system at $T=0$ lies in the strong coupling region $\alpha<<
1$, the thermal corrections drive it in the weak-coupling region where the 
condensation energy is maximum. We feel that this phenomenon gives a rather 
convincing evidence that the higher-order radiative corrections do not modify
the spontaneous generation of the magnetic condensate. So that the spontaneous  
generation of a background magnetic field is a genuine feature of the theory.

The magnetic condensate vanishes in the massless limit. However it is
concealable that in the massless theory there is the spontaneous generation of 
the dynamical mass and, at the same time, of the magnetic condensate. A complete 
treatment  of the problem can be done by using the formalism  of the composite 
operator effective potential~\cite{CJT}. 

As  a preliminary step it is important to investigate the fermionic condensate
with the thermal corrections in the massless limit.

In the P-invariant formulation of the theory, it is known~\cite{DAS} that  the 
fermion condensate $\langle \bar{\psi} \psi \rangle$ 
in presence of a constant magnetic field 
is extremely unstable. Indeed, the condensate disappears as soon as a heat bath 
is introduced. Moreover the condensate becomes nonanalytic at finite 
temperature.

On the other hand, in the P-odd formulation of QED in (2+1)-dimensions it turns 
out that~\cite{CEA} 

\begin{equation}
\label{eq27}
  \langle {\bar {\psi}} \psi \rangle
= \left\{ \begin{array}{ll}
                       \frac {eB} {2 \pi} & \mbox{$m \rightarrow 0^-$} \\
                       0                  & \mbox{$m \rightarrow 0^+$}
                       \end{array}
                       \right. 
\end{equation}
Therefore in the  massless limit and in presence of a magnetic field,
the fermionic condensate is a non-vanishing quantity only in the case 
of a negative mass term.

In our approximation it is 
straightforward to evaluate the fermion condensate at 
finite temperature. Indeed, from Eq.(\ref{eq7}) and taking into account 
that

\begin{equation}
\label{eq28}
\langle a^{\dag}_{np} a_{np}\rangle_{\beta}
=1- \langle a_{np} a^{\dag}_{np}\rangle_{\beta}=
\frac {1} { e^{\beta E_n}+1}, \nonumber 
\end{equation}
\\
\begin{equation}
\langle b_{np} b_{np}^{\dag}\rangle_{\beta}
=1- \langle b^{\dag}_{np} b_{np}\rangle_{\beta}=
\frac {1} { e^{-\beta E_n}+1},
\end{equation}
we get 

\begin{equation}
\label{eq29}
\langle \bar {\psi} \psi \rangle_{\beta}= \frac {eB} { 2 \pi} 
\frac {1} {e^{ \beta m}+1} -
\frac {m e B} {2 \pi} \sum_{n=1}^{\infty} \frac {1} {E_n} \tanh \left( \frac 
{\beta} {2} E_n \right).
\end{equation}
Equation(\ref{eq29}) holds for both positive and negative mass. The subscripts 
$\beta$ in Eqs. (\ref{eq28}) and (\ref{eq29}) stand for the thermal average 
with respect to the Hamiltonian $H_0$, Eq. (\ref{eq5}).

In the massless limit the expansion parameter is $\hat {\beta} = \beta |m|$.
This means that the massless limit coincides with the high temperature limit. A 
straightforward calculation gives:

\begin{equation}
\label{eq30}
\langle {\bar {\psi}} \psi \rangle_{\beta} = \frac {e B} {4 \pi}+ {\cal {O}} 
(\hat {\beta}).
\end{equation}
Comparing Eq. (\ref{eq30}) with Eq. (\ref{eq27}) we see that the order of the 
limits $m \rightarrow 0 $ and $ \beta \rightarrow \infty$ is not commutative.
Note that Eq. (\ref{eq30}) implies that in presence of a heat bath even with an 
infinitesimal  temperature the massless limit is symmetric:

\begin{equation}
\label{eq31}
\lim_{m \rightarrow 0^+} \langle \bar {\psi} \psi \rangle_{\beta}= 
\lim_{m \rightarrow 0^-} \langle \bar {\psi} \psi \rangle_{\beta}=
\frac {eB} {4 \pi},
\end{equation}
and it corresponds to half filled zero modes. It is remarkable that in the high 
temperature limit both the free energy and fermion condensate are accounted for 
with an effective system composed of half filled zero modes at zero temperature
.

In conclusions we have investigated (2+1)-dimensional QED coupled with P-odd 
Dirac fermions at finite temperature. We find that  the spontaneous  generation 
of a constant background magnetic field is stable towards the thermal  
fluctuations. Moreover, it turns out that  at high temperature the thermal 
fluctuations tend to increase both the condensation energy and the strength of 
the induced magnetic field. 

In addition we found that the finite temperature fermion condensate in the 
massless limit is different from zero and temperature independent. However it 
seems that, due to the non commutativity of the massless limit and the zero 
temperature limit, the fermion condensate is nonanalytic. The nonanalyticity 
vanishes if at zero temperature one assumes the rule of the symmetric massless 
limit.

Let us, finally, compare the results of the present work with 
Ref.~\cite{KANEMURA}. The authors of Ref.~\cite{KANEMURA} consider the model
studied in Ref.~\cite{HOSOTANI} at finite temperature and in the one-loop 
approximation. In the first paper of Ref.~\cite{CEA} one can found a critical 
comparison at zero temperature. At finite temperature  our results agree with 
the ones of Ref.~\cite{KANEMURA} as concern the survival of the spontaneous 
generation of the constant background field to the thermal corrections even at 
infinite temperature. Moreover we agree on the fact that the coefficient of the
term linear in the magnetic field  is accounted for by the contributions due to 
the zero modes. However, we find that the thermal corrections to the above 
mentioned coefficient are small and vanish at infinite temperature, while in 
Ref.~\cite{KANEMURA} the coefficient grows logarithmically with the 
temperature. As a consequence both the induced magnetic field and the negative 
condensation energy grow with the temperature, so that one must supply from an 
external source, via the chemical potential, the energy needed in order to 
induce the states with a nonzero magnetic field.

For these reasons we believe that the model of Ref.~\cite{KANEMURA} 
is artificial and without direct physical applications. On the other hand, 
as we have already discussed, our model is relevant for the 
dynamics of four dimensional fermions localized on the two-dimensional
space of the domain walls.


%
\newcommand{\InsertFigure}[2]{\newpage\phantom{.}\vspace*{-3cm}%
\phantom{.} \vspace*{5.truecm}
\hspace*{.5truecm} ${\cal {F}}_0 $ 
\vspace*{-8.truecm}
\begin{center}\mbox{%
\epsfig{bbllx=1.truecm,bblly=2.8truecm,bburx=16.5truecm,bbury=28.truecm,%
height=18.truecm,figure=#1}}
\end{center}\vspace*{-.1truecm}%
\parbox[t]{\hsize}{\small\baselineskip=0.5truecm\hskip0.5truecm #2}}

\InsertFigure{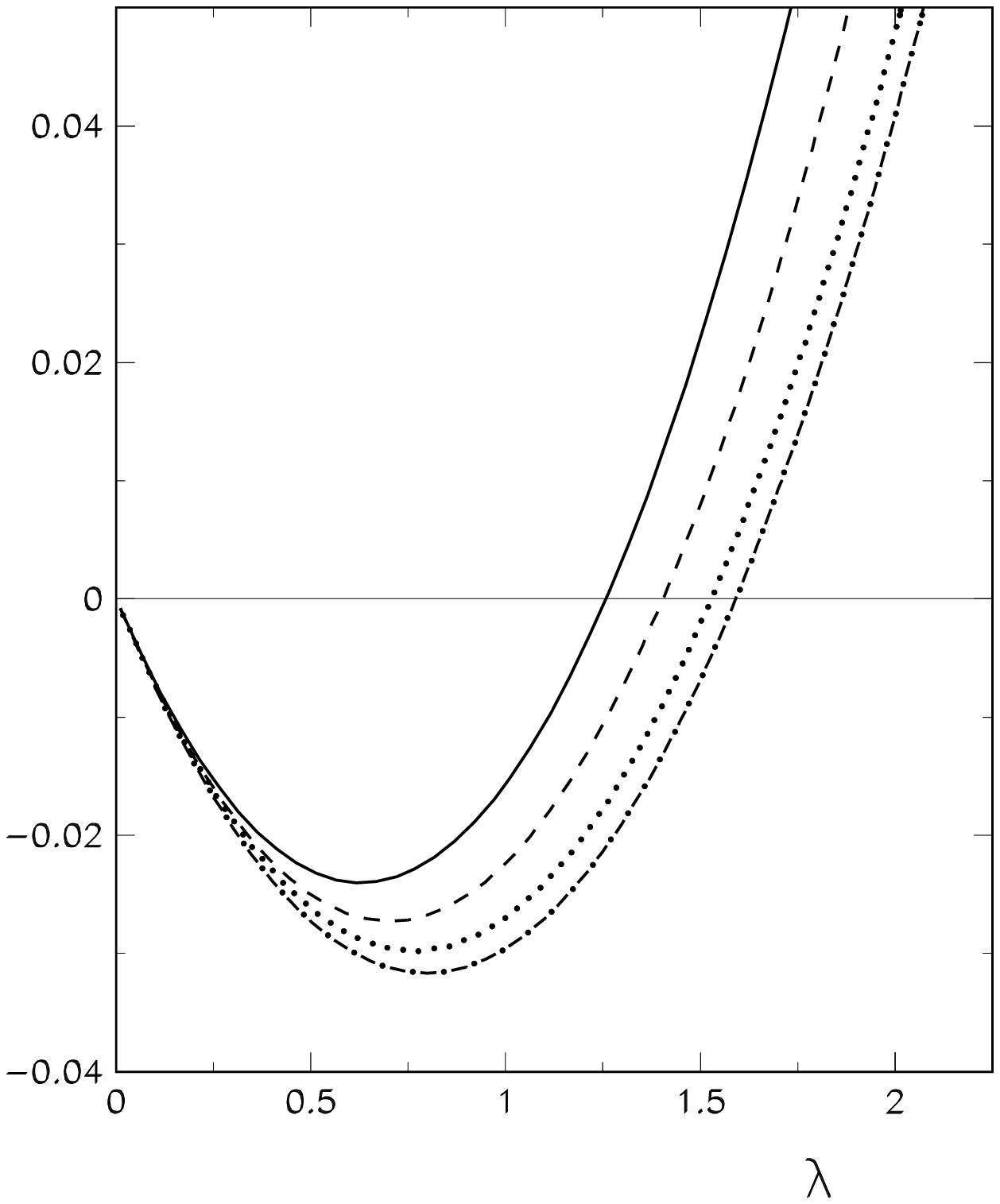}%
{FIG.~1      
                   Free energy density (in units of $|m|^3$) versus $\lambda$ 
                   for $\alpha=0.1.$ Full line  $T=0$, dashed line 
                   $\hat {T} =1$, dotted line $\hat {T} =5$, and dash-dotted 
                   line $\hat {T} \rightarrow \infty$.}
 
\InsertFigure{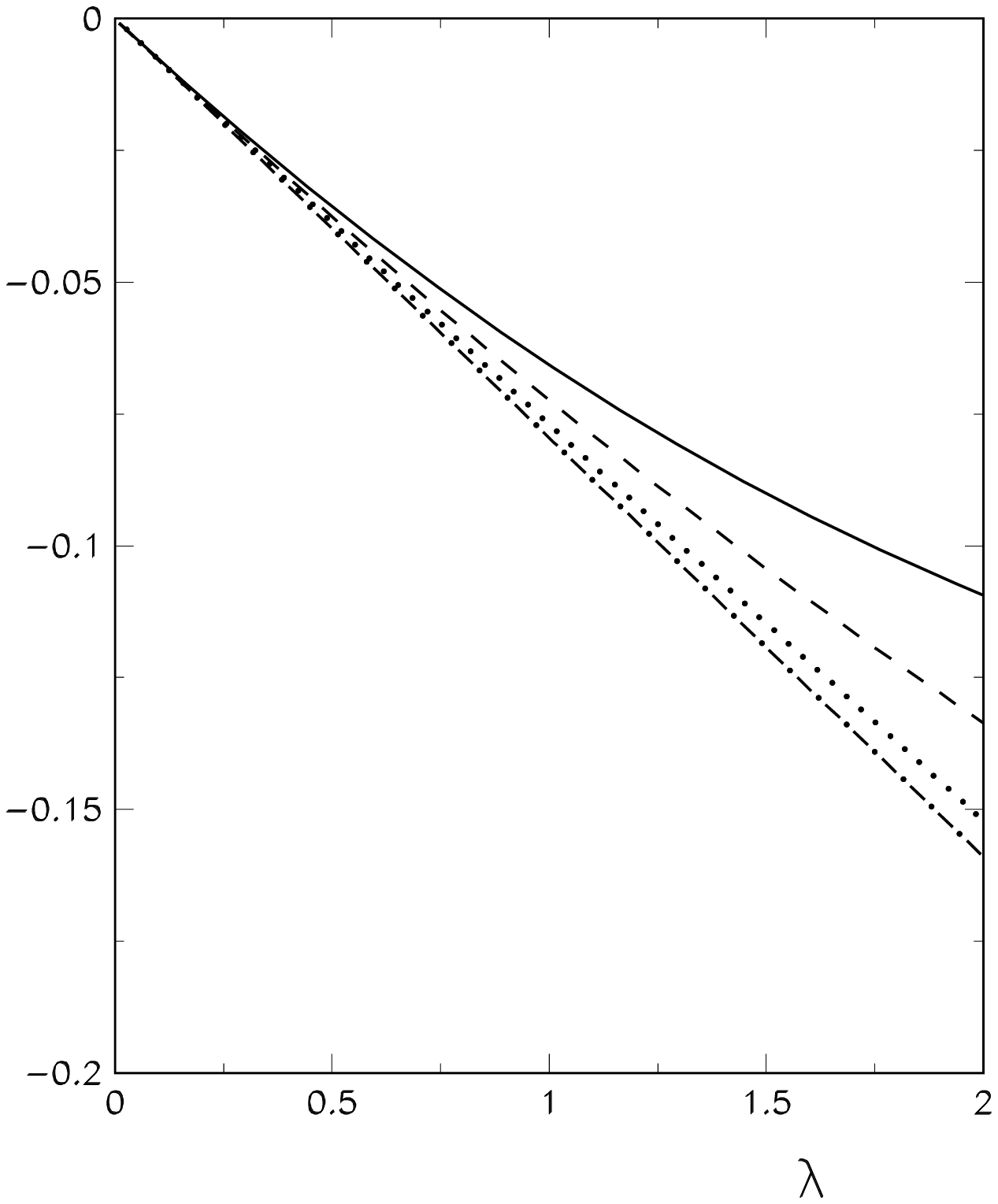}%
{FIG.~2      
                    Free energy density (in units of $|m|^3$) versus $\lambda$
                    for $\alpha=0$. Temperature values as in Fig. 1}

\end{document}